\documentclass{iaus}
\usepackage{graphicx}

\title[Chemical signature of gas-rich disc-disc mergers] 
{Chemical signature of gas-rich disc-disc mergers at high redshift}

\author[Hugo Martel et al.]   %% give here short author list %%
{Hugo Martel$^{1,2}$, Simon Richard$^{1,2}$, Chris B. Brook$^3$, 
Daisuke Kawata$^4$,
Brad K. Gibson$^3$, \and Patricia S\'anchez-Bl\'azquez$^{5,6,7}$}

\affiliation{$^1$D\'epartement de physique, de g\'enie physique et
d'optique, Universit\'e Laval, Qu\'ebec, QC, G1K 7P4, Canada \\
[\affilskip]
$^2$Centre de Recherche en Astrophysique du Qu\'ebec \\
[\affilskip]
$^3$Jeremiah Horrocks Institute for Astrophysics and Supercomputing, 
University of Central Lancashire, Preston PR1 2HE, England \\
[\affilskip]
$^4$Mullard Space Science Laboratory, University College London,
 Holmbury St Mary, Dorking RH5 6NT, England \\
[\affilskip]
$^5$Instituto de Astrof\'\i sica de Canarias, E-38200 La Laguna, 
Tenerife, Spain \\
[\affilskip]
$^6$Departamento de Astrof\'\i sica, Universidad de La Laguna, 
E-38205 La Laguna, Tenerife, Spain \\
[\affilskip]
$^7$Departamento de F\'\i sica Te\'orica, M\'odulo C15, Universidad
 Aut\'onoma de Madrid, E28049 Cantoblanco, Spain \\
}

\pubyear{2008}
\volume{xxx}  %% insert here IAU Symposium No.
\pagerange{119--126}
\jname{Title of your IAU Symposium}
\editors{A.C. Editor, B.D. Editor \& C.E. Editor, eds.}
\begin{document}

\maketitle

\begin{abstract}
We performed numerical simulations 
of mergers between gas-rich disc galaxies, which result 
in the formation of late-type galaxies. Stars formed during the merger end
up in 
a thick disc that is partially supported by velocity 
dispersion and has high $\rm[\alpha/Fe]$ ratios at all metallicities. 
Stars formed later end up in a thin, rotationally 
supported disc which has lower $\rm[\alpha/Fe]$ ratios. While 
the structural and kinematical properties of the merger remnants depend 
strongly upon the orbital parameters of the mergers, we find a clear 
chemical signature of gas-rich mergers.
\end{abstract}

\firstsection % if your document starts with a section,
\section{Introduction}

Mergers of late-type galaxies have long been
associated with the formation of
early-type galaxies (\cite[Toomre 1977]{toomre77}). 
However, numerical simulations of major 
collisions between disc galaxies 
(e.g. \cite[Naab \& Burkert 2003]{nb03}),
which lead to the formation of ellipticals, have assumed 
a low ratio gas/stars representative of low-redshift galaxies.
Going back in time, we expect to find
fewer and fewer stars, and therefore a more important
gaseous component. We have simulated mergers
of gas-rich disc galaxies, with different mass ratios and
orbital parameters (\cite[Brook et al. 2007]{brooketal07};
\cite[Richard et al. 2010]{richardetal10}).
Our main goal is to find out if there is a clear 
observational signature of gas-rich disc-disc mergers.

\begin{table}
 \begin{center}
  \caption{Initials Conditions for All Simulations}
  \label{table-initial}
  \smallskip
  \begin{tabular}{@{}lcccccc@{}}
  \hline
    Run & $M_{\rm Gal1} (M_\odot)$ &
    mass ratio & $\theta$ &
    $V_{2z}\;{\rm(km\,s^{-1}})$ & ${\bf R}_2\;{\rm(kpc)}$ 
    & $f_{\rm gas}$ \\
  \hline
 M12     & $5.0\times10^{11}$ & 2:1  &  30.0 &   0 & 0,  0, 15 & 0.94 \\
 M12orb  & $5.0\times10^{11}$ & 2:1  &  30.0 &   0 & 0, 50, 15 & 0.67 \\
 M12z    & $5.0\times10^{11}$ & 2:1  &  30.0 & 100 & 0,  0, 15 & 0.83 \\
 M1290   & $5.0\times10^{11}$ & 2:1  &  90.0 &   0 & 0,  0, 15 & 0.96 \\
 rM12    & $5.0\times10^{11}$ & 2:1  & 210.0 &   0 & 0,  0, 15 & 0.91 \\
 M11     & $2.5\times10^{11}$ & 1:1  &  30.0 &   0 & 0,  0, 15 & 0.88 \\
 M13     & $5.0\times10^{11}$ & 3:1  &  30.0 &   0 & 0,  0, 15 & 0.92 \\
 M110    & $5.0\times10^{11}$ & 10:1 &  30.0 &   0 & 0,  0, 15 & 0.85 \\
   \hline
  \end{tabular}
 \end{center}
\end{table}

\section{The Numerical Simulations}

\begin{figure}
\begin{center}
\vspace{-0.2in}
\includegraphics[width=4.0in]{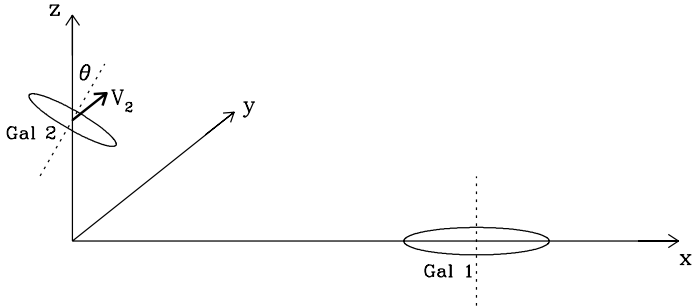}
\vspace{-0.2in}
\caption{Geometry of the initial conditions. The $Y$-axis is pointing away 
from the viewer. Gal1 is initially at rest, while Gal2 is initially
moving away from the viewer, in the $+Y$ direction. The dotted
lines indicate the rotation axes of the galaxies. 
Gal1 is in the $XY$ plane, with
the rotation axis in the $Z$-direction. The rotation axis of Gal2 is
in the $XZ$ plane, at an angle $\theta$ relative to the $Z$-axis.
Both galaxies are rotating clockwise when seen from above.
\label{fig-initial}}
\end{center}
\end{figure}

All simulations were performed using the chemo-dynamical
algorithm GCD+ (\cite[Kawata \& Gibson 2003a, 2003b]{kg03a,kg03b}). 
The initial conditions consist of
two galaxies with exponential gas discs embedded in dark matter halos.
The most massive galaxy (Gal1) has total mass 
of $5\times10^{11}M_\odot$ (except for simulation M11). 
The mass of the second galaxy (Gal2)
depends on the mass ratio chosen for the simulation.  
Fig.~\ref{fig-initial} shows the initial configuration of the system. 
Gal1 is at rest
at ${\bf R}_1=(50,0,0)\,{\rm kpc}$.
Gal2 is located at ${\bf R}_2=(0,0,15){\rm kpc}$, with
velocity ${\bf V}_2=(0,100,0){\rm km\,s^{-1}}$ (except for 
simulations M12orb and M12z).
The baryon/DM mass fraction is 17\%.
Table~\ref{table-initial} lists
the parameters used for each simulation. 
The last column lists the gas fraction at the beginning of the simulation.
In each case, the collision leaves
a very chaotic system, which then relaxes to a ``quiescent state''
where the final structure is well-established,
and the star formation rate has dropped significantly. 

\section{Results}

\begin{figure}
\begin{center}
\includegraphics[width=2in]{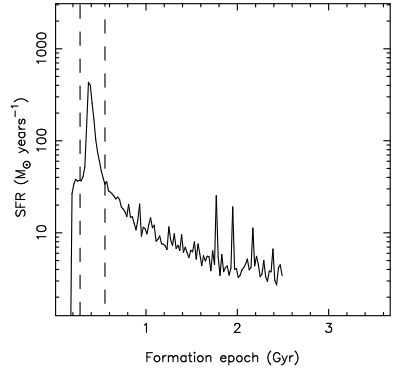}
\caption{Star formation rate versus time, for simulation M12.
The dashed lines indicate the beginning and the end of the 
starburst, respectively.
}
\label{SFR}
\end{center}
\end{figure}

We have computed the star formation rate for all simulations (Fig.~\ref{SFR}).
In all cases,
a major starburst invariably occurs during the merger.
After the starburst, the star
formation steadily drops, as the system relaxes and a disc forms.
We define two stellar populations: {\it old stars},
which are formed before and
during the starburst, and {\it young stars},
which include all stars formed after the merger, when the starburst is
completed. 

\begin{figure}
\begin{center}
\includegraphics[width=4.5in]{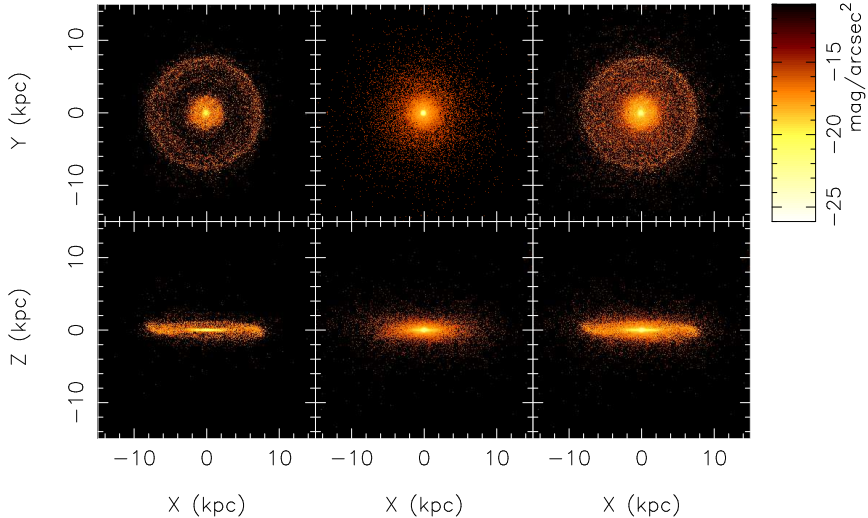}
\caption{V-band image of M12 seen face-on (top panels) and
edge-on (bottom panels), for young stars (left panels), old
stars (middle panels), and all stars (right panels). $X$, $Y$, $Z$
represent cartesian coordinates, with $Z$ along the axis of rotation.
}
\label{M12}
\end{center}
\end{figure}

Fig.~\ref{M12} show the luminosity maps of the final remnant
for simulation M12. The 
left panels show young stars, the middle panels show old stars,
and the right panels show all stars. Young and old
stars form discs that have comparable radii, but the edge-on views
(bottom panels) clearly shows that the young disc is thin, while the
old disc is thick. 
Most simulations result in the formation of a thin
disc made of young stars, and a thick disc made of old stars.
except simulations M12z and M11 which show more elliptical-like remnants.
In several cases, the merger resulted in the formation of a ring 
made of young stars. 
Simulations M1290 and
rM12 form a central bar made of both old and young stars.
From this analysis, we conclude that the structural properties of
the remnants are strongly dependent of the initial 
conditions.

\begin{figure}
\begin{center}
\includegraphics[width=4.7in]{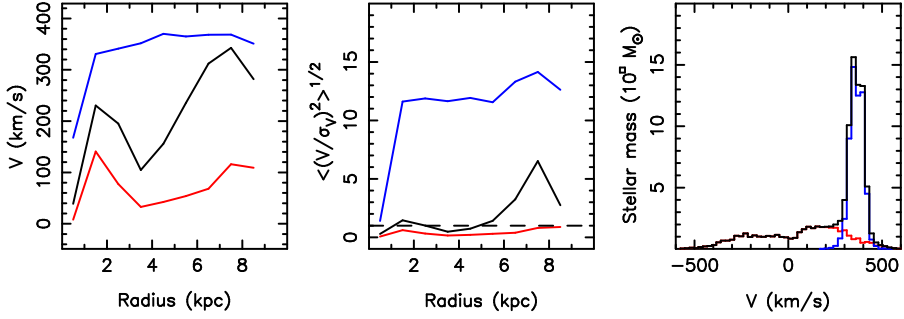}
\caption{Kinematics for simulation M12.
On each panel, red, blue, and black curves
represent old stars, young stars, and all stars, respectively.
Left: rotation velocity vs. radius. 
Middle: Rotational support versus radius. Right: histogram of stellar
mass vs. rotation velocity, with negative values
indicating counter-rotating stars.
}
\label{M12_kin}
\end{center}
\end{figure}

Fig.~\ref{M12_kin} shows the kinematical properties for simulation M12.
We calculated for each population the average rotational velocity $V$ 
and its dispersion $\sigma_V^{\phantom2}$ in radial bins.
The left panel shows the rotation curves.
In all cases, the old 
stars have a lower mean rotational velocity than the 
young stars. The middle panel shows
$\langle(V/\sigma^{\phantom2}_V)^2\rangle^{1/2}$ versus radius. 
In all cases, we find $\langle(V/\sigma^{\phantom2}_V)^2\rangle^{1/2}>1$
for young stars, indicating the presence of a
massive, rotationally supported disc. 
Most cases have
$\langle(V/\sigma^{\phantom2}_V)^2\rangle^{1/2}<1$, for old stars,
indicating the presence
of a ``hot disc'' supported by internal motions,
but simulations M1290, rM12, and M110 all have
$\langle(V/\sigma^{\phantom2}_V)^2\rangle^{1/2}>1$, indicating that
the old stars are also rotationally supported.
The right panel shows
histograms of the stellar mass versus rotation velocity.
In most cases, the young stars
are concentrated in a narrow region of the histogram,
while the old stars are spread in velocity.
The young component is counter-rotating for simulation M11,
and partly counter-rotating 
for simulation M12z. 
We conclude that
the kinematical properties of the remnants
are strongly dependent of the initial 
conditions.

\begin{figure}
\begin{center}
\includegraphics[width=3.2in]{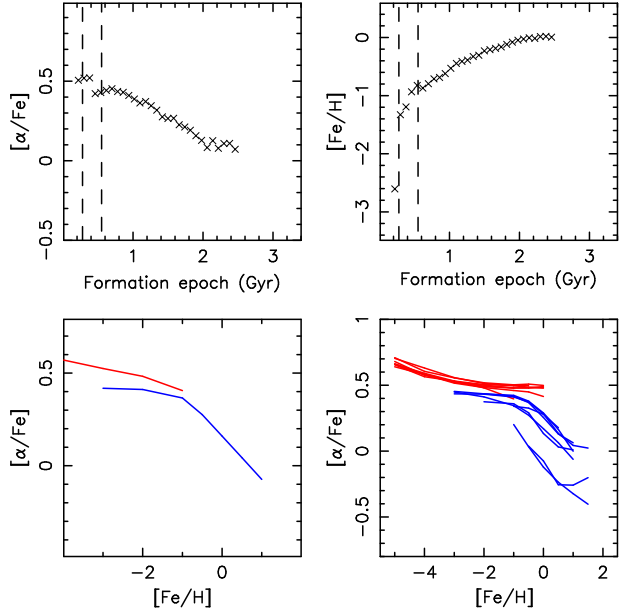}
\caption{Top panels: $\alpha$-elements abundance and 
metallicity versus formation epoch for simulation M12.
Bottom panels: $\rm[\alpha/Fe]$ versus $\rm[Fe/H]$ for old stars
(red curve) and young stars (blue curve) for simulation M12
(left) and all simulations (right).
}
\label{M12_chem}
\end{center}
\end{figure}

The chemical properties are presented in 
Fig.~\ref{M12_chem}. 
The top panels show $\rm[\alpha/Fe]$ and
$\rm[Fe/H]$ versus formation epoch for simulation M12. 
$\rm[\alpha/Fe]$ decreases with time, while
$\rm[Fe/H]$ increases
strongly before and during the merger, and keeps increasing, but more
slowly, after the merger. 
The bottom left panel shows
$\rm[\alpha/Fe]$ versus $\rm[Fe/H]$ for simulation M12. 
Old stars have larger 
$\alpha$-element abundance than young stars, up to a relatively high
metallicity ($\rm[Fe/H]\simeq-0.5$). 
$\rm[\alpha/Fe]$ decreases with
time as the metallicity increases, until it reaches the values found in the 
thin disc. This decrease of $\rm[\alpha/Fe]$ is slowed near the 
beginning of the 
collision, as the starburst leads to a large number of Type~II SNe,
that enrich the gas in $\alpha$-elements. After the starburst, Type~Ia
SNe become effective, and enrich the gas in iron.
This explains why $\rm[\alpha/Fe]$
gradually decreases after the collision. 
The bottom right panel shows $\rm[\alpha/Fe]$
versus metallicity for all 8 simulations. The results for old
stars (red curves) are nearly identical for all simulations.
Hence, the mergers leave a clear chemical
signature: the old stars, located in the
thick disc and the halo, have a ratio $\rm[\alpha/Fe]$ that remains constant 
at $\sim0.5-0.6$
with increasing $\rm[Fe/H]$ up to high metallicities ($\rm[Fe/H]=-0.5$),
while the young stars, located in the thin disc, have a lower
ratio $\rm[\alpha/Fe]$ which decreases with increasing metallicities.

%\begin{discussion}

%\end{discussion}

\end{document}